\documentclass{natureprintstylev4}
\usepackage{astjnlabbrev-nature}
\usepackage{hyperref}
\usepackage[switch]{lineno}

\usepackage{amssymb}
\usepackage{amsmath}	
\usepackage{mathtools}
\usepackage{gensymb}
\usepackage{color}
\usepackage{enumitem}
\usepackage{textcomp}

\usepackage{color}
\usepackage{epsfig}
\usepackage{color}
\usepackage{graphicx}
\usepackage{longtable}
\usepackage{hyperref}
\usepackage{soul}

\usepackage[switch]{lineno}

\usepackage[labelsep=endash]{caption}

\usepackage{marvosym}

\newcommand{\citep}{\cite}
\newcommand{\citet}{\cite}

\title{A high-velocity star recently ejected by an intermediate-mass black hole in M15}

\author{Yang Huang$^{1,3,7,\dagger,\textsuperscript{\Letter}}$, Qingzheng Li$^{2,1,\dagger}$, Jifeng Liu$^{3,1,7,\textsuperscript{\Letter}}$,  Xiaobo Dong$^{2,\textsuperscript{\Letter}}$, Huawei Zhang$^{5,6,\textsuperscript{\Letter}}$, Youjun Lu$^{1,4}$, Cuihua Du$^{1}$}

\begin{document}
\setstcolor{red}
\maketitle

\begin{affiliations}
\item School of Astronomy and Space Science, University of Chinese Academy of Sciences, Beijing 100049,  People's Republic of China;\\
\item Yunnan Observatories, Chinese Academy of Sciences, Kunming 650011, China;\\
\item New Cornerstone Science Laboratory, National Astronomical Observatories, Chinese Academy of Sciences, Beijing 100012, China;\\
\item National Astronomical Observatories, Chinese Academy of Sciences, Beijing 100012, China;\\
\item Department of Astronomy, School of Physics, Peking University, Beijing 100871, China;\\
\item Kavli Institute for Astronomy and Astrophysics, Peking University, Beijing 100871, China;\\
\item Institute for Frontiers in Astronomy and Astrophysics, Beijing Normal University, Beijing, 102206, China;\\
$\dagger$ These authors contributed equally;\\
$\textsuperscript{\Letter}$ Corresponding authors:  jfliu@nao.cas.cn; huangyang@ucas.ac.cn; xbdong@ynao.ac.cn; zhanghw@pku.edu.cn.
\end{affiliations}
\\

\section*{Abstract}
The existence of intermediate-mass black holes (IMBHs) is crucial for understanding various astrophysical phenomena, yet their existence remains elusive, except for the LIGO-Virgo detection. 
We report the discovery of a high-velocity star J0731+3717, whose backward trajectory  about $21$\,Myr ago intersects that of globular cluster M15 within the cluster tidal radius.
Both its metallicity [Fe/H] and its alpha-to-iron abundance ratio [$\alpha$/Fe] are consistent with those of M15. 
Furthermore, its location falls right on the fiducial sequence of the cluster M15 on the color-absolute magnitude diagram, suggesting similar ages.
These support that J0731+3717 is originally associated with M15 at a confidence level of ``seven nines".
We find that such a high-velocity star ($V_{\rm ej} = 548^{+6}_{-5}$\,km\,s$^{-1}$) was  most likely tidally ejected from as close as one astronomical unit to the center of M15, confirming an IMBH 
($\ge 100 M_{\odot}$ with a credibility of 98\%) as the exclusive nature of the central unseen mass proposed previously. 

\textbf{Keywords:} Intermediate-mass black hole, Hills mechanism, hypervelocity star, globular cluster, large-scale survey

\section*{Introduction}
Intermediate-mass black holes (IMBHs) in the mass range of $10^2$--$10^5$ solar mass ($M_{\odot}$) may fill the gap between the BHs formed as stellar remnants and the supermassive BHs (SMBHs) found in the centers of galaxies\cite{2020ARA&A..58..257G}.
Except for LIGO-Virgo detection\cite{2020PhRvL.125j1102A, 2023PhRvX..13d1039A}, their existence, however, is still uncertain despite extensive searching efforts.
Discovering IMBHs and characterizing their mass function in this range are thus of great interest for many reasons\cite{2020ARA&A..58..257G}.

Globular clusters (GCs), dense and massive dynamical systems, have long been expected as promising places to harbor IMBHs.
Both theory and numerical simulations suggest that IMBHs can form either through the repeated mergers of stellar black holes (``slow mode")\cite{2002MNRAS.330..232C}, remnants of massive stars that sink to the center, or through the explosion of a very massive star resulting from runaway mergers of stars during an early phase of cluster core collapse (``fast mode")\cite{2002ApJ...576..899P}.
The ``slow mode" is a competitive scenario to explain the LIGO-Virgo detected IMBHs\cite{2020PhRvL.125j1102A, 2023PhRvX..13d1039A}, yielded from mergers of binary BHs.

Extensive efforts have indeed found a large central unseen mass in some globular clusters by velocity dispersion\cite{2002AJ....124.3270G, 2008ApJ...676.1008N} or pulsar timing measurements\cite{2017Natur.542..203K}, but such a mass can be either an IMBH or a cluster of stellar remnants within a few thousand astronomical unit (AU)\cite{2006ApJ...641..852V, 2019MNRAS.488.5340B, 2017MNRAS.471..857F}.
To exclude the latter possibility, we need to limit the central mass to a much smaller volume. 
One approach is to search for hyper/high-velocity stars ejected from globular clusters.
If they exist, they are largely linked to the tidal interaction between an IMBH and a binary system for a close encounter typically within one AU.

\begin{table*}
\caption{The measured parameters of J0731+3717 and M15}
\begin{center}
\resizebox{\linewidth}{!}{
\begin{tabular}{lcccc}
\hline
Parameter&J0731+3717 & M15&Units\\
\hline
RA (J2000)& 07:31:27.26 & 21:29:58.33 & hh:mm:ss.ss \\
Dec (J2000)&$+$37:17:04.3    & $+$12:10:01.2&   dd:mm:ss.s\\
{\it Gaia} DR3 source\_id & 898707303799931392 & n.a.&n.a.\\
{\it Gaia} DR3 Proper motion $\mu_{\alpha}\rm{cos}\,\delta$ &$47.963 \pm 0.041$& $-0.659 \pm 0.024$& mas\,yr$^{-1}$\\
{\it Gaia} DR3 Proper motion $\mu_{\delta}$ &$ -83.666 \pm 0.035$ &  $-3.803 \pm 0.024$& mas\,yr$^{-1}$\\
{\it Gaia} DR3 Parallax&$0.733 \pm 0.042$&  $0.097 \pm 0.010$  & mas\\
{\it Gaia} DR3 $G$-band magnitude&$15.814 \pm 0.003$& n.a.&mag\\
{\it Gaia} DR3 $G_{\rm BP} - G_{\rm RP}$&$0.765 \pm 0.007$& n.a.&mag\\
SDSS $u$-band magnitude&$17.095 \pm 0.008 $& n.a.&mag\\
SDSS $g$-band magnitude&$16.194 \pm 0.004 $& n.a.&mag\\
SDSS $r$-band magnitude&$15.832 \pm 0.004$& n.a.&mag\\
SDSS $i$-band magnitude&$15.693 \pm 0.004$& n.a.&mag\\
SDSS $z$-band magnitude&$15.643 \pm 0.007$& n.a.&mag\\
Distance&$1295.2 \pm 13.1$ & $10709.0 \pm 95.5$&pc\\
$E (B - V)$&$0.047$& 0.08 &mag\\
Heliocentric radial velocities HRV &$196.68 \pm 6.97$ & $-107.0 \pm 0.2$ & km\,s$^{-1}$\\
Effective temperature $T_{\rm eff}$&$6062.0 \pm 157.0$ & n.a. &K\\
Surface gravity log\,$g$&$4.02 \pm 0.29$& n.a. &dex\\
Metallicity [Fe/H]&$-2.23 \pm 0.13$ & $-2.33 \pm 0.02$ (stat.) $\pm 0.10$ (syst.) &dex\\
$\alpha$-element to iron ratio [$\alpha$/Fe]&$+0.24 \pm 0.07$ & $0.24 \pm 0.03$ &dex\\
%Galactocentric distance $r_{\rm GC}$&$9392.1^{+70.0}_{-58.3}$ & $10750.7^{+66.5}_{-66.5}$ &pc\\
Total velocity $V_{\rm GSR}$& $418.71^{+6.54}_{-5.95}$ & $111.85^{+1.48}_{-1.35}$ &km\,s$^{-1}$\\
Age&$13.00^{+1.75}_{-2.00}$& $12.5 \pm 0.3$ &Gyr\\
Mass&$0.69^{+0.02}_{-0.01}$ & $4.94 \times 10^{5}$ &$M_\odot$\\
\hline
\end{tabular}}
\end{center}
\label{table:Table1}
\end{table*}

\section*{Results and Discussion}
\textbf{A high-velocity star ejected from globular cluster M15}

To discover high-velocity stars ejected from globular clusters, backward orbital integrations are carried for 934 high-velocity ($V_{\rm GSR} \ge 400$\,km\,s$^{-1}$) halo stars in the searching volume within 5\,kpc from the Sun\cite{2023AJ....166...12L} and 145 Galactic globular clusters\cite{2010arXiv1012.3224H, 2021MNRAS.505.5957B, 2021MNRAS.505.5978V}.
Both trajectories of stars and globular clusters are traced back to 250 Myr ago (in about one orbital period at solar position) using a common-adopted model of steady-state Galactic potential\cite{2015ApJS..216...29B}.
The closest distance for each pair of high-velocity star and globular cluster is calculated from their backward trajectories.
Amongst the hundred thousand pairs, only J0731+3717 has closest distance smaller than the tidal radius of M15, making it a rare candidate of cluster ejected high-velocity star (More technical details can be found in Supplementary Materials A to F).

Table\,1 summarizes the information of J0731+3717 and M15.
J0731+3717  is a high-velocity star with $V_{\rm GSR} = 419_{-6}^{+6}$\,km\,s$^{-1}$ at a heliocentric distance of $1.295 \pm 0.013$\,kpc.
M15 is a 12.5 billion year old globular cluster\cite{2016ApJ...827....2V} located at the constellation Pegasus with a heliocentric distance of $10.71 \pm 0.10$\,kpc\cite{2021MNRAS.505.5957B} and a mass of $5 \times 10^5 M_{\odot}$\cite{2020MNRAS.492.3859D}.
M15 is believed to host an intermediate-mass black hole (IMBH) of 1700-3200\,$M_{\odot}$ based on velocity dispersion measurements\cite{2002AJ....124.3270G, 2003AJ....125..376G}, albeit with debates\cite{2006ApJ...641..852V, 2003ApJ...582L..21B, 2003ApJ...595..187M}, especially the $3\sigma$ upper limit of $980$\,M$_{\odot}$ placed by the non-detection from ultra-deep radio observations at the cluster core\cite{2012ApJ...750L..27S}.
While J0731+3717 is currently 11.5\,kpc away from M15, their backward trajectories intersect with each other 21\,Myr ago with a relative velocity of  $548_{-5}^{+6}$\,km\,s$^{-1}$ and a closest distance of $58$\,pc (Figure\,1a), smaller than the cluster tidal radius of $132$\,pc.
The intersection locations and their uncertainties in 3D space are estimated with one million Monte Carlo (MC) trajectory simulations to be $\Delta X = 20_{-135}^{+122}$\,pc, $\Delta Y = 19_{-104}^{+95}$\,pc and $\Delta Z = 51_{-32}^{+32}$\,pc, as shown in Figures 1b-d.

In addition to the orbital connection, J0731+3717 exhibits rare chemical fingerprints consistent with those of M15.
The SEGUE spectrum\cite{2009AJ....137.4377Y} clearly shows J0731+3717 is a very metal-poor late F-type star, as plotted in Figure 2a, with  effective temperature $T_{\rm eff} = 6062 \pm 157$\,K, metallicity [Fe/H] = $-2.23 \pm 0.13$ and alpha-to-iron abundance ratio [$\alpha$/Fe] =\,$+0.24 \pm 0.07$ (see Supplementary Materials~E).
The old globular cluster M15 has well-measured [Fe/H]\,=\,$-2.33 \pm 0.10$ and [$\alpha$/Fe]\,=\,$+0.24 \pm 0.03$ \cite{2009A&A...508..695C, 2016A&A...590A...9D}.
Their chemical parameters are consistent with each other within errors, and are both located in a region with very few stars on the [Fe/H]--[$\alpha$/Fe] plane.
Only $7.5 \times 10^{-3}$ of all halo stars within the 5\,kpc searching volume having reliable abundances are located in such a region shown in Figure 2b.
The rare chemical similarity implies J0731+3717 was originally associated with the cluster, in line with the suggestion of orbital analysis.

The association of J0731+3717 with M15 can be further supported by their similar ages as derived from isochrone fitting.
Broad-band photometric measurements for M15 and J0731+3717 are adopted from the Sloan Digital Sky Survey (SDSS) Galactic globular and open clusters project\cite{2008ApJS..179..326A}  and SDSS DR12\cite{2015ApJS..219...12A}, respectively.
Both their absolute magnitudes and colors are corrected for extinction values along their separate lines of sight.
As shown in Figure\,2c, J0731+3717 falls right on the cluster fiducial sequence of M15; subsequently, its isochrone age as estimated from Bayesian approach is $13.00^{+1.75}_{-2.00}$ Gyr, almost identical to the age obtained for M15 (see Supplementary Materials F and G). 
In comparison, 37\% of the field halo stars with chemical abundances similar to M15 (defined by the box in Figure 2b) actually deviate from its fiducial sequence by more than 0.05\,mag (the maximal error) in color direction, implying significantly different ages.

It is extremely unlikely for the association of J0731+3717 and M15 to be by pure chance, given the probability for random association, chemical and age similarities.
To quantitively determine the probability of random high-velocity halo stars to encounter M15, we run MC simulations to generate about one million high-velocity halo stars with $V_{\rm GSR} \ge 400$\,km\,s$^{-1}$ in such a searching volume (see Supplementary Materials H).
%given the probability for random high-velocity halo stars in the 5\,kpc searching volume to intersect and share similar chemical and age similarities with cluster M15.
%The later two probabilities have been already estimated.
%First, the fraction of high-velocity stars with $V_{\rm GSR} \ge 400$\,km\,s$^{-1}$ is only 1.93\% (964,630).
Only 12 high-velocity stars have close encounters with M15 within its tidal radius in the past 250\,Myr.
This means the probability of unphysical orbital encounter between J0731+3717 and M15 is $1.2 \times 10^{-5}$.
Considering the orbital link, together with the chemical and age similarities, one high-velocity halo star in our searching volume is coincidently linked to M15 by a pure chance of only $5.89 \times 10^{-8}$.
In other words, J0731+3717 is a true former member of M15 at a confidence level of  ``seven nines".

\begin{figure*}
 \centerline{\includegraphics[scale=1.2,angle=0]{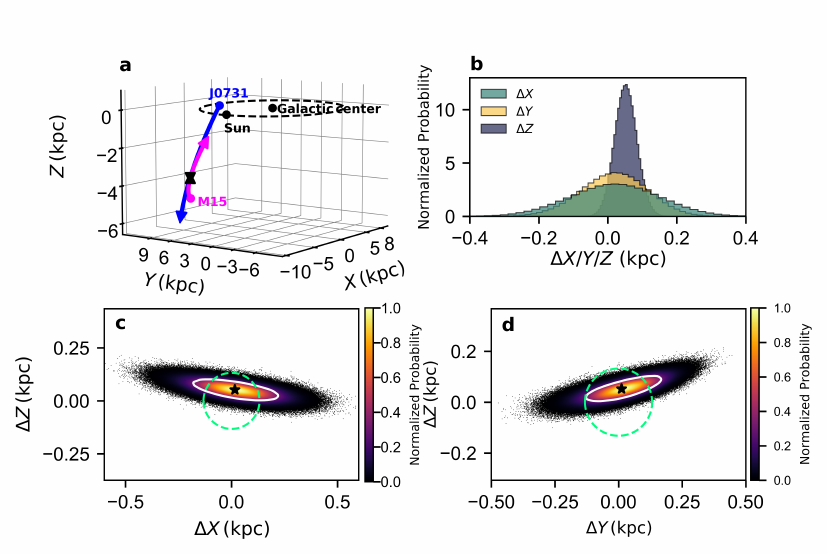}}
\caption{\textbf{Backward orbital analysis of J0731+3717 and M15.} \textbf{a,} 3D representation of the backward orbits of J0731+3717 and the globular cluster M15.
The blue and magenta lines with arrows mark the backward orbits of J0731+3717 and M15, respectively.
The triangle and inverted triangle mark the encounter positions, happened 21 Myr ago, for M15 and J0731+3717.
The black dots represent the locations of the Galactic center and the Sun as labelled.
The solar circle ($R = 8.178$\,kpc) is shown by the dashed black line.
\textbf{b,}  
The distributions of the closest distance between J0731 and 
the center of M15 along $X$ (green), $Y$ (yellow) and $Z$ (purple) at encounter yielded by one million MC trajectory simulation (see Supplementary Materials B).
\textbf{c/d,}  Density of the relative positions (J0731+3717 with respect to M15) at encounter, again yielded by  the one million MC trajectory calculations, on (\textbf{c}) $\Delta X$--$\Delta Z$ and (\textbf{d}) $\Delta Y$--$\Delta Z$  planes.
The normalized number density is color coded, as shown by the right colorbar.
%The color represents the number density as indicated by the normalized probability in the right side.
The white contours mark $1\sigma$ confidence region.
The green dashed circles represent the size of tidal radius of M15.
The black star represent the relative location at closest distance derived directly using the observational parameters listed in Table\,1.}
\label{fig:plane_countour}
\end{figure*}

\begin{figure*}
 \centerline{\includegraphics[scale=0.9,angle=0]{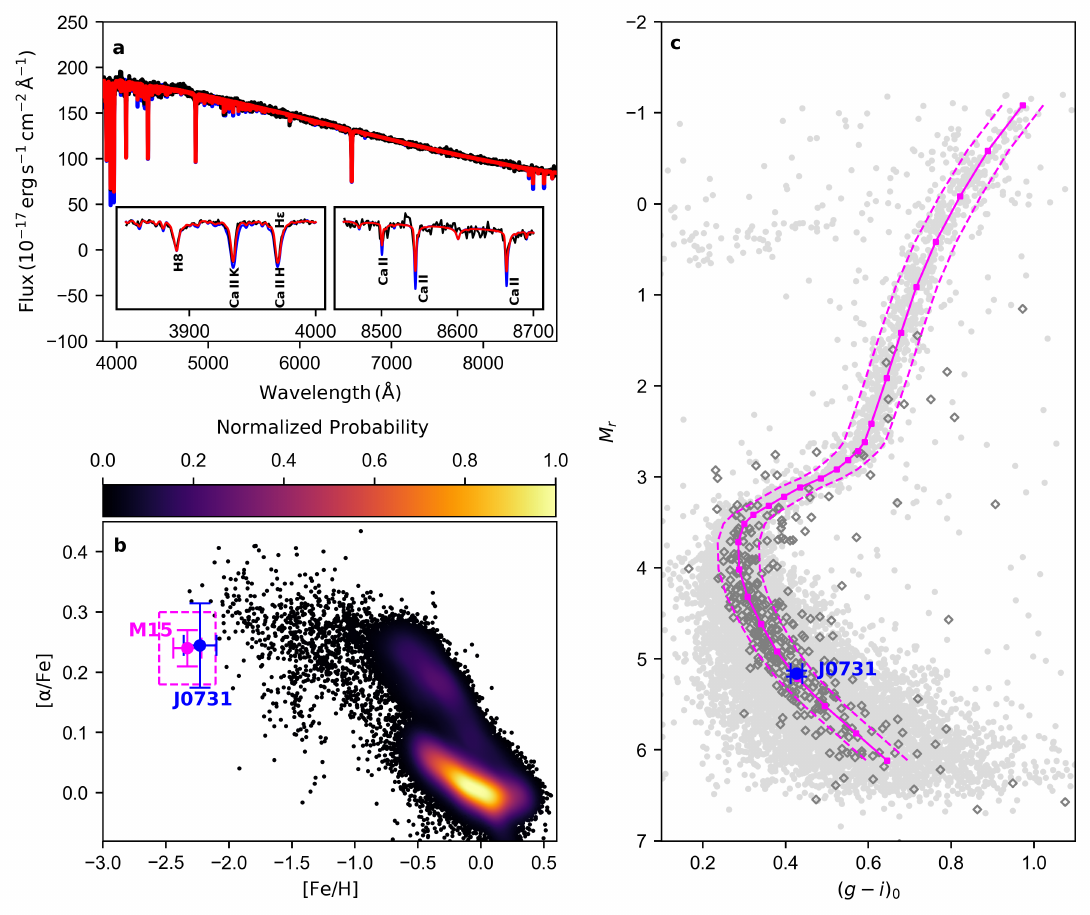}}
\caption{\textbf{Optical spectrum, chemical properties and color-absolute magnitude diagram of J0731+3717 and M15.}
\textbf{a,} Optical spectrum (in black) of J0731+3717 from the SEGUE survey. 
Two synthetical spectra (degraded to SEGUE spectral resolution) are overplotted for comparisons (see Supplementary Materials E).
The red one has stellar parameters ($T_{\rm eff} = 6100$\,K, log\,$g = 4.0$, [Fe/H]\,=\,$-2.0$ and [$\alpha$/Fe]\,=+0.20) similar to those of J0731+3717, while the blue one is 0.5\,dex richer in metallicity (i.e. [Fe/H]\,=\,$-1.5$) with the other parameters unchanged.
The insets zoom in the regions of Ca {\sc ii} H ($\lambda$3968) and K ($\lambda$3933) lines, and Ca {\sc ii} triplet lines at $\lambda\lambda$8498, 8542, 8662.
\textbf{b,} [Fe/H]--[$\alpha$/Fe] diagram for J0731+3717 (blue dot) and globular cluster M15 (magenta dot). 
The dashed magenta box marks the region within two times of measurement uncertainties of [Fe/H] and [$\alpha$/Fe] for M15.
For comparison purpose, the background shows the density of APOGEE targeted stars with reliable determinations of [Fe/H] and [$\alpha$/Fe] (see Supplementary Materials H).
\textbf{c,} $M_r$ versus $(g - i)_0$ diagram of globular cluster M15 and J0731+3717. 
The background gray dots are photometric observations of M15 from the Sloan Digital Sky Survey Galactic globular and open clusters project\cite{2008ApJS..179..326A}, by adopting a cluster distance of 10.71\,kpc and an $E (B-V)$ value of 0.08 (Table\,1).
The magenta squares denote the cluster fiducial sequence that is derived from the background gray dots by ref.\cite{2008ApJS..179..326A}.
The dashed magenta lines are shifted from the fiducial sequence by $\pm 0.05$\,mag in $(g-i)_0$.
The diamonds are field halo stars from the existing spectroscopic surveys with chemical fingerprints (defined by the magenta box shown in panel b) similar to M15 (see Supplementary Materials H).}
\label{fig:plane_countour}
\end{figure*}

\begin{figure*}
 \centerline{\includegraphics[scale=0.95,angle=0]{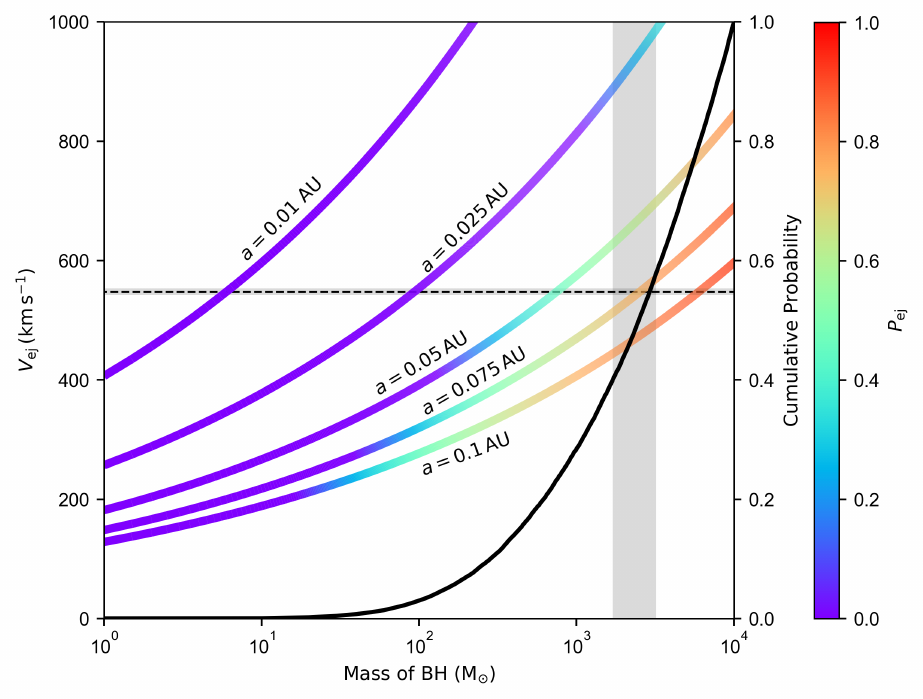}}
\caption{\textbf{Ejection velocities predicted by Hills mechanism.}
The lines represent the most probable ejection velocities, calculated by Equation~S7 in Supplementary Materials~I, as a function of black hole mass under different binary separation ranging from 0.01 to 0.1\,AU (from left to right).
In the calculation, the binary is assumed to contain two J0731+3717 like stars, each with a mass of $0.69 M_{\odot}$.
The color of lines indicates the ejection probability derived from Equation~S9 in Supplementary Materials~I with the colorbar shown on the right side.
Here the closest distance that the binary can approach to the black hole is set to 0.5~AU.
The horizontal dashed line with a shaded $1\sigma$ uncertainty indicates the reported ejection velocity ($548^{+6}_{-5}$\,km\,s$^{-1}$) of J0731+3717.
The vertical shaded region represents the  mass of black hole (1700-3200\,$M_{\odot}$) of M15 dynamically derived from velocity dispersion measurements\cite{2002AJ....124.3270G, 2003AJ....125..376G}.
The black line represents the cumulative probability of the ejection of 0731+3717-like stars at different masses of black holes, calculated through Monte Carlo simulations (see Supplementary Materials I).}
\label{fig:plane_countour}
\end{figure*}

\textbf{Interpretation as Hills ejection via IMBH}

An energetic ejection mechanism is required to kick-off J0731+3717 from M15 with an ejection velocity up to $548^{+6}_{-5}$\,km\,s$^{-1}$; in comparison,  the cluster central escape velocity is only 62\,km\,s$^{-1}$ (the escape velocity at the cluster half-mass radius is 27 \,km\,s$^{-1}$; ref.\cite{2002ApJ...568L..23G}).
The Hills mechanism\cite{1988Natur.331..687H} invokes three-body exchange interactions between stellar binary and super-massive BHs to eject hypervelocity stars from the Galactic center; such a mechanism can naturally eject high-velocity stars from the IMBHs in the center of globular clusters. 
If this mechanism is at work in this case, the ejection velocity would constrain the mass of central BH in M15 to be 726\,$M_{\odot}$ for a binary separation $a = 0.05$\,AU and $5804$\,$M_{\odot}$ for $a = 0.10$\,AU.
A stellar-mass BH smaller than $100 M_{\odot}$ can also eject a star up to around 550\,km\,s$^{-1}$ if $a \le0.025$\,AU, but the probability is only few percent  based on our MC simulations of two hundred million Hills ejections (see Supplementary Materials~I and Figure~3).
Thus, the ejection of J0731+3717 from M15 requires an IMBH ($\ge 100 M_{\odot}$ with a credibility of 98\%) at the center of the cluster; the encounters mostly occurred (93\%) within 2 AU from the cluster center.
This confirms that the previously claimed aggregated mass of more than a few thousand solar mass\cite{2002AJ....124.3270G, 2003AJ....125..376G, 2012ApJ...750L..27S} is indeed an IMBH rather than a cluster of stellar remnants\cite{2003ApJ...582L..21B, 2003ApJ...595..187M}.

\textbf{Excluding alternative explanations}

A competing mechanism to eject high-velocity stars (even up to 2000\,km\,s$^{-1}$) in globular cluster is the single star-binary interaction involving compact objects. 
Observations reveal the presence of a binary neutron star possibly as remnant from these encounters in M15 \cite{1991Natur.349..220P}.
However, the ejection rate, as derived by a recent comprehensive Monte Carlo N-body simulation\cite{2023ApJ...953...19C}, for producing J0731+3717 like high-velocity stars remains remarkably low, approximately $4 \times 10^{-8}$\,yr$^{-1}$ at present-day, which is three orders of magnitude lower than that of aforementioned IMBH--binary encounters (see Supplementary Materials~J). 
The same simulation shows that no J0731+3717 like high-velocity stars with $V_{\rm GSR} \ge 400$\,km\,$^{-1}$ were kicked from M15 through this channel in the past 10 Gyr (let alone the past 250\,Myr).

High ejection velocity can be obtained through several alternative scenarios.
First, a star can be kicked through normal single star-binary interaction\cite{2012ApJ...751..133P} without compact objects involved or exchange collision with a massive star\cite{2009MNRAS.396..570G} (similar to Hills mechanism but for massive stars), but the typical ejection velocity is within 200\,km\,s$^{-1}$.
To kick-off a J0731+3717 like star with ejection velocity above 550\,km\,s$^{-1}$, interaction between a very massive star (50--100\,$M_{\odot}$) and a hard massive binary is required\cite{2009MNRAS.396..570G}, which is impossible to happen in such an old cluster M15 most recently.
Second, a star can be ejected by the core-collapse supernova explosion of its former massive companion in a binary scenario\cite{1961BAN....15..265B}.
However, this is unlikely occurred recently in the old cluster M15 and the maximum kick velocity is largely within 300-400\,km\,s$^{-1}$\cite{2020MNRAS.497.5344E}.
At present, only the type Ia supernovae from the white dwarf plus helium star (hot subdwarf) or the dynamically driven double-degenerate double-detonation\cite{2009A&A...508L..27W, 2018ApJ...865...15S}
channels can eject the surviving helium star or white dwarf with velocity even up to 1000\,km\,s$^{-1}$ like US\,708\cite{2015Sci...347.1126G} or like D6-1 to D6-3\cite{2018ApJ...865...15S}.
One possible such ejection associated with the GC was recently reported in the nearby galaxy NGC~5353\cite{2024ApJ...968L...6B}.
The late F-type nature of J0731+3717 certainly rules out the possibility of a fast helium star or white dwarf ejected by the type Ia supernovae explosion.
Third, a star can be stripped from the cluster when the latter experiences tidal shock from interactions with giant molecular clouds\cite{2006MNRAS.371..793G}, Galactic disk\cite{1997ApJ...474..223G}, spiral arm\cite{2007MNRAS.376..809G} or perigalactic passages\cite{1997ApJ...474..223G}.
Apparently,  J0731+3717 is not such a case since the ejection position (at $-3.6$\,kpc below the Galactic disk plane; see Figure 1a) is far away from the Galactic disk, known giant molecular clouds or the last pericenter of M15.
In summary, the above alternative ejection mechanisms are not viable to kick-off J0731+3717 from M15.

\section*{Conclusion}
Our discovery of J0731+3717 ejected by an IMBH in M15 thus proves that the existence of IMBHs can be disclosed by high-velocity stars ejected from clusters via Hills mechanism, unambiguously as compared to previous velocity dispersion measurements\cite{2002AJ....124.3270G, 2006ApJ...641..852V, 2003ApJ...582L..21B, 2003ApJ...595..187M}.
Such a method can be applied to find more cases like J0731+3717 in our Galaxy.
Simulations of the 145 globular clusters in the past 14 Gyr led to around 500 J0731+3717 like high-velocity stars ejected into the current 5\,kpc searching volume (see Supplementary Materials J, although only 50 of them were ejected in the past 250\,Myr that can be traced back to their cluster origin under current measurement uncertainties.
With the increasing power of ongoing {\it Gaia} and large-scale spectroscopic surveys, we expect to discover dozens of cases within the 5\,kpc volume and ten times more within a 10\,kpc volume, which should shed light on the understanding of the evolutionary path from stellar-mass BHs to SMBHs. 

\section*{Methods}
Detailed methods are available in the Supplementary data.

\section*{Data Availability}
All data used in this study is publicly available. 
This work made use of data from the European Space Agency (ESA) mission Gaia (\url{https://www.cosmos.esa.int/gaia}),
processed by the Gaia Data Processing and Analysis Consortium (DPAC, \url{https://www.cosmos.esa.int/web/gaia/dpac/consortium}).
The stellar parameters of J0731+3717 is  available from the SDSS-II/III SEGUE archive at \url{https://data.sdss.org/sas/dr12/sdss/sspp/ssppOut-dr12.fits}. 
The SDSS photometric and spectroscopic data can be found at \url{http://skyserver.sdss.org/dr12/en/tools/search/radial.aspx}.
The photometric catalog of M15 is available from \url{http://das.sdss.org/va/osuPhot/v1_0/}.
The stellar isochrones can be found at \url{http://stellar.dartmouth.edu/models}.
The data supporting the plots in this paper and other findings of this study are available from the corresponding authors upon reasonable request.

\section*{Code Availability}
We use standard data analysis tools in the Python environments. Specifically, the orbital analysis is carried out with Python package Gala and galpy, which is publicly available at \url{http://gala.adrian.pw/en/v1.6.1/getting_started.html} and \url{www.galpy.org}.

\section*{Supplementary Data}
Supplementary data are available at {\it NSR} online.

\section*{Acknowledgements}
This work presents results from the European Space Agency (ESA) space mission Gaia. Gaia data are being processed by the Gaia Data Processing and Analysis Consortium (DPAC). Funding for the DPAC is provided by national institutions, in particular the institutions participating in the Gaia MultiLateral Agreement (MLA). The Gaia mission website is {\url{https://www.cosmos.esa.int/gaia}}. The Gaia archive website is {\url{https://archives.esac.esa.int/gaia}}.

\section*{Funding}
This work is supported by the Strategic Priority Research Program of the Chinese Academy of Sciences (XDB0550103).
Y.H. acknowledges the National Science Foundation of China (1242200163, 11903027 and 11833006), and  the National Key RD Program of China (2023YFA1608303 and 2019YFA0405503). 
JFL acknowledges support the National Science Foundation of China (11988101 and 11933004), and support from the New Cornerstone Science Foundation through the New Cornerstone Investigator Program and the XPLORER PRIZE.
XB.D. acknowledges the support from the National Science Foundation of China (12373013).
H.W.Z. acknowledges the National Science Foundation of China (11973001, 12090040 and 12090044), and the National Key RD Program of China (2019YFA0405504).

\section*{Author Contributions}
Y.H. led this project and wrote the paper;
Q.Z.L. contributed to the sample preparation, systematic search, data analysis, and wrote the manuscript together with Y.H.;
J.F.L. contributed to the interpretation of the results and writing the text;
H.W.Z. contributed to the project planning, discussions, and text revisions;
X.B.D. contributed to the project planning, discussions, and text revisions;
Y.J.L. contributed to the interpretation and revisions of the text;
C.H.D. contributed to the interpretation and revisions of the text.

%-------------------------
% References
%-------------------------
\section*{Reference}
\bibliographystyle{scibull}
\bibliography{cas-refs}

%\begin{addendum} 

%\item[Competing interests statement]
%The authors declare no competing interests.

%\item[Author Information] 
%Correspondence and requests for materials should be addressed to Y.H. (huangyang@ucas.ac.cn), J.F.L. (jfliu@nao.cas.cn), X.B.D. (xbdong@ynao.ac.cn),  and H.W.Z. (zhanghw@pku.edu.cn).

%\end{addendum}

\end{document}

\onecolumn
\setcounter{section}{0}
\renewcommand{\thesection}{S\arabic{section}}
\setcounter{equation}{0}
\renewcommand{\theequation}{S\arabic{equation}}
\setcounter{figure}{0}
\renewcommand{\thefigure}{S\arabic{figure}}
\setcounter{table}{0}
\renewcommand{\thetable}{S\arabic{table}}
\maketitle
\begin{affiliations}
\item School of Astronomy and Space Science, University of Chinese Academy of Sciences, Beijing 100049,  People's Republic of China;\\
\item Yunnan Observatories, Chinese Academy of Sciences, Kunming 650011, China;\\
\item New Cornerstone Science Laboratory, National Astronomical Observatories, Chinese Academy of Sciences, Beijing 100012, China;\\
\item National Astronomical Observatories, Chinese Academy of Sciences, Beijing 100012, China;\\
\item Department of Astronomy, School of Physics, Peking University, Beijing 100871, China;\\
\item Kavli Institute for Astronomy and Astrophysics, Peking University, Beijing 100871, China;\\
\item Institute for Frontiers in Astronomy and Astrophysics, Beijing Normal University, Beijing, 102206, China;\\
$\dagger$ These authors contributed equally;\\
$\textsuperscript{\Letter}$ Corresponding authors:  jfliu@nao.cas.cn; huangyang@ucas.ac.cn; xbdong@ynao.ac.cn; zhanghw@pku.edu.cn.
\end{affiliations}
\\

\clearpage
\section*{Supplementary Materials}
\appendix

\section{Coordinate systems.}
We adopt two sets of coordinate systems: (1) a right-handed Cartesian coordinate system ($X$, $Y$, $Z$), with $X$ towards the direction opposite of the Sun, $Y$ pointing to the direction of Galactic rotation, and $Z$ in the direction of north Galactic pole; (2) a Galactocentric cylindrical system ($R$, $\phi$, $Z$), with $R$ the projected Galactocentric distance increasing radially outwards, $\phi$ the azimuthal angle towards the direction of the Galactic rotation, and $Z$ the same as that in the Cartesian system.
The Sun is set to at ($R_0$, $Z_{\odot}$) = ($8.178$, $0.025$)\,kpc (ref.\citep{GRAVITY19, Bland-Hawthorn16}) and the circular velocity at
the solar position is fixed to $V_c (R_0) = 220$\,km\,s$^{-1}$ (ref.\cite{Bovy15}).
The solar motions with respect to the local standard of rest are set to  $(U_{\odot}, V_{\odot}, W_{\odot})=(7.01, 10.13, 4.95)$\,km\,s$^{-1}$ (ref.\cite{Huang15}).

\section{Systematic Search.}
By combing datasets of the RAVE\,DR5 (ref.\citep{Kunder17}), SDSS\,DR12 (ref.\citep{Alam15}), LAMOST\,DR8 (ref.\citep{Luo15}), APOGEE\,DR16 (ref.\citep{Jonsson20}), GALAH\,DR2\citep{Buder18}, and {\it Gaia}\,DR3 (ref.\citep{Gaia2023A}), a large sample of 943 high-velocity halo stars with total velocity $V_{\rm GSR} \ge 400$\,km\,s$^{-1}$ are constructed\citep{Li+23}.
The distances of those stars are first estimated from the parallax measurements provided by {\it Gaia}\,DR3 using a Bayesian approach\citep{Li+23}.
The distance estimates are further improved by adding constraints from SDSS optical colors and chemical information (i.e., metallicity and [$\alpha$/Fe] from spectroscopic surveys) into the Bayesian analysis, which will be introduced in the latter section.
Their 3D velocities are then calculated from the measured distances, proper motions again from {\it Gaia}\,DR3 and heliocentric radial velocities (HRV) provided by the aforementioned spectroscopic surveys, except for the SDSS survey, whose HRV is carefully re-determined by the LAMOST stellar parameter pipeline at Peking University (LSP3; ref.\citep{Xiang15}) utilizing specific template of similar atmospheric parameters for each star (see technique details in latter section).
Moreover, they have metallicities and alpha-to-iron abundance ratios well measured from their spectra.
We perform a systematic search for high-velocity stars ejected from globular clusters.
To do so, backward orbital integrations are carried for those 943 high-velocity halo stars and 145 Galactic globular clusters with well measured metallicities, distances, proper motions and HRVs (refs.\citep{Harris10}$^{,}$\citep{VB21}$^{,}$\citep{BV21}). 
For the orbital analysis, we adopt the python package {\tt Gala}\citep{Price-Whelan17} with the Galactic potential setting to {\tt MilkyWayPotential}\citep{Bovy15}, which contains four components: a nucleus, a bulge, a disk, and a dark matter halo.
%we adopt the python package {\tt Gala}\citep{Price-Whelan17} with the Galactic potential setting to {\tt MilkyWayPotential}\citep{Bovy15}, which contains four components: a nucleus, 
In this study, we only concern recent ejections during the past 250\,Myr since the dynamical effects (e.g., the dynamical friction\citep{Chandrasekhar43}), that globular cluster experienced, are hard to be precisely considered in the backward orbital calculations. 
The trajectories are therefore traced back in time of 250\,Myr with a time resolution of 1\,kyr.
To discover cluster ejected high-velocity candidates, the trajectories of each high-velocity star and globular clusters are carefully investigated by sorting the ratios between the closest orbital distances $d_{\rm min}$ and the tidal radii $r_{\rm t}$ of the clusters which can be found at \url{https://people.smp.uq.edu.au/HolgerBaumgardt/globular/parameter.html}.
The final candidates are required to have the ratios ($d_{\min}$/$r_{\rm t}$) smaller than one and metallicity differences between the high-velocity stars and the clusters smaller than 0.2\,dex.
During this search, we identified J0731+3717 (the only one of 135,430 trajectory pairs), a high-velocity star has a close encounter with M15 within its tidal radius about 21\,Myr ago.
%\st{To improve the integration precision, the distances of J0731+3717 are re-determined by adding constrains of SDSS $ugriz$ photometric observations to the Bayesian  analysis and its HRV is carefully re-determined by a specific metal-poor template (see details in latter sections).
%The trajectories of J0731+3717 and M15 are then refined with updated distance and HRV of J0731+3717.}
The closest orbital distance is $58.0_{-58.0}^{+117.4}$\,pc, which is smaller than the tidal radius ($r_{\rm t} = 132.1$\,pc) of M15 (see Extended Data Table\,1 for top 5 clusters sorted by ratios of $d_{\min}$/$r_{\rm t}$).  
The encounter occurred where J0731+3717 is located at ($X$, $Y$, $Z$)\,=\,($-5.91^{+0.13}_{-0.14}, 7.19^{+0.10}_{-0.10}, -3.61^{+0.04}_{-0.04}$)\,kpc with ($V_{X}$, $V_{Y}$, $V_{Z}$)\,=\,($-205.16^{+6.07}_{-6.68}, -305.94^{+6.23}_{-6.85}, 171.19^{+3.06}_{-3.37}$)\,km\,s$^{-1}$, and M15 is located at ($X$, $Y$, $Z$)\,=\,($-5.93^{+0.03}_{-0.03}, 7.17^{+0.09}_{-0.09}, -3.66^{+0.03}_{-0.03}$)\,kpc with ($V_{X}$, $V_{Y}$, $V_{Z}$)\,=\,($56.18^{+2.15}_{-1.95},103.13^{+1.26}_{-1.26},-82.10^{+1.38}_{-1.30}$)\,km\,s$^{-1}$.
At the closest approach, J0731+3717 had a relative velocity of  $541.10^{+5.96}_{-5.42}$\,km\,s$^{-1}$ with respect to M15 (see Extended Data Figure 1).
The upper and lower uncertainties of these reported parameters from our orbital analysis come from 16 and 84 per cent percentiles of the probability distribution function (PDF) yielded by one million MC trajectory calculations.
 During the calculations, the astrometric parameters (except the parallax/distance) of J0731+3717 are assumed to multivariate Gaussian distributions with the correlation coefficients (taken from {\it Gaia} DR3) considered. The distance of J0731+3717 is assumed to follow the posterior PDF derived from our Bayesian analysis. The HRV of J0731+3717 is uncorrelated to astometric parameters and thus assumed normally distributed.
 For M15, the uncertainties in positions, distance, proper motions and HRV are taken from refs\citep{VB21}$^{,}$\citep{BV21}. and are all assumed to follow normal distribution.

We note that the last cross with the disk is happened 2.6 Myr ago with the intersect positions at $(X_{p}, Y_{p}) = (-9.1, 0.9)$\,kpc, far away from the Galactic center. 
This result clearly rules out the possibility of ejecting J0737+3717 from the supermassive BH at the Galactic center.
In addition to the above 250\,Myr orbital integrations, the whole past 14\,Gyr (the age of universe) backward orbits of J0731+3717 and globular clusters, as well as the best-known dwarf galaxies (parameters are adopted from ref.\citep{F18}), are investigated. 
No other orbital links are found. The whole backward orbits of J0731+3717 are also provided as an online supplementary data.

\section{Orbital deflections.}
During the backward orbital integrations, the deflections caused by encounters with field stars are ignored.
We thus here evaluate this effect on orbital analysis.
The deflection angle is defined as\citep{BT08}:
\begin{equation}
\theta_{\rm defl} = 2\tan^{-1}(b_{90}/b),
\end{equation}
where $b$ is the distance of the close approach, and $b_{90}$ is the $90^{\circ}$ deflection radius that is given by:
\begin{equation}
b_{90} = \frac{G(m_1 + m_2)}{V_0^2},
\end{equation}
where $m_1$ and $m_2$ are the masses of two stars having close meet, $V_0$ is the relative speed.
By assuming two solar-like stars with $V_0 = 300$\,km\,s$^{-1}$, the deflection angle is only about 0.039 arcsec for a 1\,pc encounter.
The Sun is expected to experience 1\,pc encounter at a rate of $19.7 \pm 2.2$ Myr$^{-1}$\citep{BJ18}.
During our orbital calculations with a total integration time of 250\,Myr, the maximum deflection angle for solar-like stars at solar position is within 3.5 arcmin, which is much smaller than the tidal radius of M15 (42.4 arcmin for $r_{\rm t} = 132$\,pc at 10.7\,kpc).
In our case,  the deflection angles of these halo stars are even much smaller, given their much lower spatial number density.
Specifically, for J0731+3717 with typical $V_0 = 420$\,km\,s$^{-1}$, the deflection angle is smaller than 8.4 arcsec (0.44\,pc at the distance of M15), even assuming a high encounter rate at solar position.
This deflection angle is not only much smaller than the tidal radius of M15 or uncertainty of meet distance (see above section) but also significantly less than its half-light radius (one arcmin from ref.\citep{Harris10}).
Thus, the orbital deflection has minor effects on our backward orbital analysis.
%However, it is crucial to consider this effect when determining the exact position of an ejection inside the globular cluster -- a highly dense system of stars.

\section{Orbital analysis with alternative assumptions.}
To show the robustness of our backward orbital reconstructions, we repeat the whole analysis by adopting alternative assumptions.
In total,  six tests are performed.
In each test, only one assumption is changed and other parameters remain the same.
First, we change the Galactic potential to {\tt BovyMWPotential2014}\citep{Bovy15} consisting three components: a bulge, a disk, and a dark matter halo.
In the second test, alternative values of solar motions respect to the local standard of rest ($U_{\odot}, V_{\odot}, W_{\odot}$)\,=\,($11.10, 12.24, 7.25$)\,km\,s$^{-1}$ are adopted\citep{S10}.
In the third test, we use the circular velocity at the solar position $V_{c} (R_0) = 234.04$\,km\,s$^{-1}$ (ref.\citep{Zhou23}).
 In the fourth test, the Galactocentric distance of the sun $R_0$ is changed to 8.34\,kpc\citep{Reid14}.
For the fifth test, M15 is modelled as a moving potential with a Plummer profile, calculated by the {\tt MovingObjectPotential} function implemented in {\tt galpy} (ref.\citep{Bovy15}). The size and mass of M15 is taken from ref.\citep{BV21}.
Finally, we consider the effects from the potential of the Large Magellanic Cloud (LMC), by modelling it as a Plummer profile with a total mass of $1.38 \times 10^{11}$\,M$_{\odot}$ and a scale radius of 17.14 kpc (ref.\citep{Erkal19}).
The sky positions, distances, proper motions and HRV of LMC are adopted from ref.\citep{Patel2020}.
The moving potential of LMC is agian calculated by {\tt MovingObjectPotential}.
We also include the dynamical friction on the LMC from the Milky Way by the {\tt ChandrasekharDynamicalFrictionForce} function of {\tt galpy}.
The closest encounter distances as found by the above tests are all within 60\,pc (see Table~S2 ), smaller than the tidal radius ($r_t = 132.1$\,pc) of M15.
The backward time of the meets are all around 21\,Myr, in great consistency with our default result.
The comprehensive tests show that the impact of alternative assumptions on the backward orbital analysis is negligible.

\section{SEGUE spectra and parameters of J0731+3717.}
Two optical spectra of J0731+3717 have been obtained by the SEGUE survey\citep{Yanny09} on 2000 November 29 and 2005 March 17, with a signal-to-noise-ratio (SNR) of 49.4 and 55.8, respectively.
The spectra have covered the full optical range (3800-9200\AA) with a resolving power of around 2000.
Stellar parameters and heliocentric radial velocity (HRV) are derived from the observed spectra by the SEGUE Stellar Parameter Pipeline (SSPP)\citep{Lee08a}.
The typical precisions are 157\,K, 0.29\,dex, 0.13\,dex, 0.07\,dex and 5\,km\,s$^{-1}$ for effective temperature $T_{\rm eff}$, surface gravity log\,$g$, metallicity [Fe/H], alpha-to-iron abundance ratio [$\alpha$/Fe], and heliocentric radial velocity (HRV), respectively\cite{Lee08a, Lee08b}.
Two groups of stellar parameters are derived by SSPP from the two visits and they are consistent with each other very well, showing the robustness of the pipeline.
We here adopt the stellar parameters and HRV (Table\,1) measured from the spectrum with highest SNR, i.e. the one observed on 2005 March 17.
The spectrum is shown in Fig.\,2a.
Clearly, the SEGUE spectrum indicates that J0731+3717 is a real very metal-poor star with [Fe/H] down to $-2$ and even lower, directly shown by the comparisons with the synthetical spectra adopted from G{\"o}ttingen spectral library\citep{Husser13} (see Figure 2b).

Given the metal-poor nature of J0731+3717, we re-estimate its HRV and uncertainty by using a specific metal-poor template.
To do so, we adopted the LSP3 that is developed to derive HRVs from LAMOST/SEGUE like low resolution spectra using cross-correlating technique with ELODIE library\citep{PS01} as template (degraded to SEGUE resolution).
We note the SEGUE spectrum (observed on 2005 March 17) of J0731+3717 was corrected for a systematic offset of $+7.3$\,km\,s$^{-1}$ due to wavelength calibrations (ref.\citep{Adelman-McCarthy08}; \url{https://www.sdss3.org/dr9/algorithms/wavelength.php}).
For the template, we choose BD+023375 ($T_{\rm eff} = 5944$\,K, log\,$g = 3.97$ and [Fe/H]\,=\,$-2.29$; ref.\citep{PS01}) star included in the ELODIE library, whose atmospheric parameters are very close to these of J0731+3717.
We then run LSP3 to the SEGUE spectrum of J0731+3717 and find its HRV of $196.68 \pm 6.97$\,km\,s$^{-1}$.
The $1\sigma$ error is properly estimated using metal-poor stars with multiple observations in LAMOST.
This uncertainty is slightly larger than the typical one (5\,km\,s$^{-1}$) for normal metal-rich FGK stars but in great consistent with the independent check from metal-poor globular clusters (ref.\citep{Lee08b}).
As mentioned in previous systematic search section, the HRVs of other high-velocity stars from SDSS DR12 are also re-determined by LSP3 in the above manner.

%In summary, the orbital deflection has minor effects on our backward orbital analysis.
%Here we evaluate the effect of the orbital deflection due to encounter with background stars 
%In solar neighborhood, solar-like stars could be deflected by up to 0.28 arcsec per Myr due to encounters with the background stars (see next section).
%In this degree, it is hard to trace back the orbits of stars ejected longer than 200\,Myr ago, whose orbits could deviate the original ones by as large as one arcmin, larger than the tidal radii of most of known globular clusters.

\section{Mass, age and improved distance of J0731+3717.}
We determine the mass, age and improved distance of J0731+3717 by the common used Bayesian approach.
In this approach, the PDF of the age, mass and absolute magnitudes is expressed as,
\begin{equation}
f(\tau, m, M_{\lambda}) = NP(\tau, m)\mathcal{L}(\tau, m),
\end{equation}
where $\lambda = u,g,r,i,z$ and  $N$ is a normalization parameter that ensures $\iint f(\tau, m) d\tau dm =1$. 
For $P(\tau, m)$, a uniform prior and a Salpeter initial mass function\citep{Salpeter55} are assumed for age and mass, respectively.
The likelihood function $\mathcal{L}$ is obtained by,
\begin{equation}
\mathcal{L} = \prod_{i = 1}^{n}\frac{1}{\sqrt{2\pi}\sigma_i} \times \exp(-\chi_i^2/2), 
\end{equation}
where
\begin{equation}
\chi_i^2 = \left(\frac{O_i - M_i (\tau, m)}{\sigma_i}\right)^2.
\end{equation}
Here $O_i$ represents the observational constraints from the intrinsic color $(g-i)_0$, absolute magnitudes $M_{\lambda}$ and metallicity [Fe/H], $M_i$ represents the model values from the stellar isochrones, taken from the Dartmouth Stellar Evolution Program (DSEP)\citep{Dotter08}, at a give $\tau$ and $M$.
The total number of observed parameters is $n$, and $\sigma_i$ is the uncertainty of the $i$th observed parameter.
The intrinsic color $(g-i)_0$ of J0731+3717 is from SDSS photometry after correcting for the dust reddening.
The five SDSS band absolute magnitudes are given by combination of SDSS photometry with reddening corrected and the distance of J0731+3717 measured from {\it Gaia} parallax.
Here the value of the reddening is taken from the SFD map\citep{SFD} (after correction of 14\% systematics\citep{Yuan13}) and the extinction coefficients are adopted from ref.\citep{Yuan13}.
The metallicity [Fe/H] is derived from the SEGUE spectra.
The uncertainties of these observational constraints (as listed in Table\,1) are well considered in the estimation.
For the DSEP isochrones, a constant value $+0.20$ dex of [$\alpha$/Fe], close to that measured for J0731+3717, is adopted.

The resulted PDFs yield a mass of $0.69^{+0.02}_{-0.01}$\,M$_{\odot}$, an age of $13.00^{+1.75}_{-2.00}$\,Gyr for J0731+3717 and a weighted mean distance modulus of $10.56\pm0.02$ (corresponding to $d = 1295.2 \pm 13.1$\,pc) from the five SDSS bands.
 We note this method is also adopted to improve the distance estimate of other high-velocity stars in previous section of systematic search.
In Figure~S2, we compare J0731+3717 to the DSEP isochrones on the $M_r$--$(g-i)_0$ diagram to show the robustness of the resulted parameters.
The derived age of J0731+3717 agrees very well with that determined for M15.

\section{Color-absolute magnitude diagram of M15.}
We adopt the photometric measurements from the Sloan Digital Sky Survey Galactic globular and open clusters project\citep{An08} to construct the color-absolute diagram of M15.
This project presents precise photometry for 17 Galactic globular clusters and open clusters by using the DAOPHOT/ALLFRAME procedure\cite{Stetson87, Stetson94}
that are developed for crowded fields.
We select M15 member stars from the catalog yielded by the project using the following criteria: 1) stars are within 6 arcmin from the center of M15; 2) stars in $gri$-bands are required with DAOPHOT parameters: $|$sharp$|<1$ and $\chi < 1.5 + 4.5 \times 10^{-0.4 (m-16.0)}$; 3) stars have photometric uncertainties smaller than 0.2\,mag in $gri$-bands.
In total, over ten thousand stars are left from the above cuts.
By adopting a cluster distance of 10.71\,kpc (ref.\citep{BV21}) and a dust reddening of $E (B-V) = 0.08$ (ref.\citep{Sandage13}), the color-absolute magnitude is constructed for M15 (see grey dots in Figure\,2c).
We further convert the cluster fiducial sequence (the locus of the number density peaks, taken from ref.\citep{An08}) on $r$ versus $(g-i)$ plane to that on $M_r$ versus $(g-i)_0$ plane, by adopting the cluster distance and reddening values.
The results are shown in Figure\,2c as the magenta squares.

\section{Confidence level of the ejection of J0731+3717 from M15.}
We calculate the possibility that one high-velocity halo star (in the 5\,kpc searching volume) is linked to M15 by pure chance.
First, we generate 50 million stars randomly distributed in a local volume of $(5 {\rm kpc})^3$ from the Sun (similar to the searching volume of high-velocity stars\citep{Li+23}) following velocity distributions of the local halo stars:
\begin{equation}
f (v_R, v_{\phi}, v_Z) = k\exp\left(-\frac{v_R^2}{2\sigma_R^2}-\frac{(v_{\phi} - \overline{v_{\phi}})^2}{2\sigma_{\phi}^2}-\frac{v_Z^2}{2\sigma_Z^2}\right),
\end{equation}
where $k = \frac{1}{(2\pi)^{3/2}\sigma_{R} \sigma_{\phi} \sigma_{Z}}$, $\overline{v_{\phi}} = +35.53$\,km\,s$^{-1}$, $\sigma_R = 150.57$\,km\,s$^{-1}$, $\sigma_{\phi} = 115.67$\,km\,s$^{-1}$ and $\sigma_Z = 86.67$\,km\,s$^{-1}$ (ref.\citep{Anguiano20}).
In total, around one million stars (964,630) are found at the high-velocity tail with $V_{\rm GSR} \ge 400$\,km\,s$^{-1}$.
We thus perform backward orbital analysis for those high-velocity halo stars and the globular cluster M15.
The integration settings are the same as those detailed in systematic search section.
During the calculations, only 12 mock high-velocity halo stars have the chances to meet M15 within its tidal radius ($r_{\rm t} = 132.1$\,pc).
If requiring the close meet distance smaller than 60\,pc (like J0731+3717), only 3 mock stars are left. 
The result indicates that the high-velocity halo stars in our searching volume coincidently insect with M15 by a pure chance of $1.2 \times 10^{-5}$ ($P_{\rm orbit}$).
By an empirical cut on $V_{\phi}$--[Fe/H] diagram (Figure~S3), a total of 97,464 halo stars within the 5\,kpc searching volume are found from the existing large-scale spectroscopic surveys:  SDSS\,DR12, LAMOST\,DR8, APOGEE\,DR16, and GALAH\,DR2.
This means only 0.02 star could encounter with M15 within its tidal radius; but we indeed find one: the star J0731+3717.   

Second, we estimate the possibility that stars possess chemical fingerprints similar to that of M15.
To do so, a total of 1,930,135 stars (58,594 halo stars based on the aforementioned cut on $V_{\phi}$--[Fe/H] diagram) with reliable determinations of [Fe/H] and [$\alpha$/Fe] (requiring spectral SNR greater than 30) are selected from the existing large-scale spectroscopic surveys.
Similar to the selection of high-velocity stars in ref.\citep{Li+23}, all stars are within a 5\,kpc searching volume by requiring parallax greater than 0.2\,mas, parallax measurement error better than 20\% and RUWE smaller than 1.4.
Amongst these stars, 442 stars are found with [Fe/H] and [$\alpha$/Fe] close to these of M15 within two times observational uncertainties of the cluster (see magenta box at Figure\,2b).
The values of [Fe/H] and [$\alpha$/Fe] and their uncertainties of M15 are listed in Table\,1.
Therefore, 0.75\% halo stars (442/58,594; $P_{\rm chem}$) in our searching volume have similar chemical pattern on [Fe/H]--[$\alpha$/Fe] as these of M15.

Finally, we check the age differences for those halo stars with chemical pattern similar to that of M15.
The distances from {\it Gaia} DR3 parallax measurements\citep{Li+23} and SDSS $gri$ photometric observations\citep{Alam15} are cross-matched to those 442 aforementioned stars.
By requiring photometric uncertainties smaller than 0.05 mag, 379 stars are left and shown in Figure\,2c (grey diamonds).
62.8\% of them (238/379; $P_{\rm CMD}$) fall into the region within the two dashed magenta lines in Figure\,2c and thus they have similar age as that of M15.

Overall, there is only $5.89\times10^{-8}$ ($P_{\rm orbit} \times P_{\rm chem} \times P_{\rm CMD}$) chance that J0731+3717 is not physically linked to M15. We thus conclude that J0731+3717 is ejected from M15 at a confidence level of ``seven nines".

\section{Black hole mass of M15.}
By applying Hills mechanism to globular cluster, the most probable ejection velocity for a star is (ref.\citep{Bromley06}),
\begin{equation}
v_{\rm ej} \approx 460\left(\frac{a}{0.1 {\rm AU}}\right)^{-1/2}\left(\frac{m}{2 M_{\odot}}\right)^{1/3}\left(\frac{M}{10^3 M_{\odot}}\right)^{1/6} {{\rm km\,s}^{-1}},
\end{equation}
where $a$ is binary semimajor axis, $m$ is the total mass of the binary and $M$ is the mass of the massive black hole in the globular cluster.
According to the momentum conservation, the ejection speeds for the primary and secondary are,
\begin{equation}
v_1 = v_{\rm ej} \left(\frac{2m_2}{m}\right)^{1/2}, {\rm and}\ v_2 = v_{\rm ej} \left(\frac{2m_1}{m}\right)^{1/2},
\end{equation}
respectively. $m_1$ and $m_2$ is the mass of primary and secondary.
Following ref.\citep{Hills88}, the possibility of an ejection is expressed as,
\begin{equation}
P_{\rm ej} = 1 - D/175,
\end{equation}
where $D$ is a dimensionless quantity,
\begin{equation}
D = \left(\frac{r_{\rm close}} {a}\right) \left[\frac{2M}{10^6(m_1 + m_2)}\right]^{-1/3}.
\end{equation}
Here $r_{\rm close}$ represents the closest distance that the binary can approach to the black hole of the cluster.
$P_{\rm ej} \equiv 0$ when $D > 175$.

As shown in the Fig.~S2, the maximum mass of `undead' star in the old cluster M15 is no greater than 1\,$M_{\odot}$.
Even considering white dwarf with a mass close to Chandrasekhar limit as the companion, the mass ratio of the progenitor binary of J0731+3717 is no greater than 2.
The mass ratio $q$ is then assumed to uniformly distribute from 0 to 2.
The mass of $m_1$ is equal to $qm_2$, where $m_2$ denotes the mass of J0731+3717.
%\st{For the binary semimajor axis, we adopt the log-normal distribution derived from the field G-dwarf stars}\st{, but with a maximum cut of 2\,AU.
%Soft binaries with large separations ($> 2$\,AU) are assumed to be destroyed due to the dynamic evolution of cluster}.
For the binary semimajor axis, we adopt the present-day distribution (with age up to 13.6\,Gyr; \url{https://cmc.ciera.northwestern.edu/home/}) for main-sequence binaries from the Cluster Monte Carlo (CMC) {\it N}-body simulations\citep{Kremer20}.
Amongst the 148 independent $N$-body simulations, the model {\tt N16-RV0.5-RG8-Z0.1} was chosen since its present-day properties (e.g., a central velocity dispersion of 9.25\,km\,s$^{-1}$ and a half-mass radius of 3.9 pc at 13.6\,Gyr) are very similar to these of M15 (a central velocity dispersion of 13.1\,km\,s$^{-1}$ and a half-mass radius of 3.66\,pc; ref.\cite{BH18}, the last update can also found at \url{https://people.smp.uq.edu.au/HolgerBaumgardt/globular/parameter.html}). 
%(with age up to 12\,Gyr; \url{https://zenodo.org/records/4042966}) for main-sequence binaries from the DRAGON simulations (ref.\citep{Shu21}), which is the first one-million-particle direct {\it N}-body simulations of globular clusters. There are two distributions ({\tt D1-R7-IMF93} and {\tt D2-R7-IMF01}) generated from different assumptions of initial mass function. Both of them are tried in our estimate.
As same as ref.\citep{Bromley06}, $r_{\rm close}$ is assumed to uniformly distribute between 0.1 and 700\,AU.
Finally, considering above assumptions and the observational constraints from the ejection velocity and stellar mass, as well as their uncertainties, of J0731+3717, we generate over 200 million progenitor binaries in the MC simulations to derive the mass distribution function of the black hole.
By requiring $P_{\rm ej} > 0$ and BH mass smaller than 10,000 $M_{\odot}$ ($3\sigma$ upper limit of the central dynamic mass within M15 found by previous studies\cite{Gerssen02, Gerssen03}), we find that the black hole mass of M15 is greater than 100\,$M_{\odot}$ with a credibility of 97.784\%.
Moreover, the closest distance $r_{\rm close}$ between the binary and black hole is largely (93.017\%) within 2 AU, as the ejection probability decreases rapidly with increasing $r_{\rm close}$ (See Fig~3 and Fig.~S4 for comparison).
%regardless of the two distributions of binary semimajor axis is considered.
Our comprehensive analysis strongly suggests that J0731+3717 is ejected by an IMBH hosted by M15.

Finally, we remark that we can not fully rule out close encounters happened in some rare systems, for example, a single star scattered by binary IMBHs\citep{FB19}. 
These systems are of particular interesting for various astrophysical studies but obviously far beyond the current knowledge.
%\st{or binary neutron stars}\citep{CR23}

\section{Estimate the number of J0731+3717 like star in the searching volume of current and future surveys.}
The current searching volume is limited by the {\it Gaia} DR3 parallax measurement that is only accurate for estimating stars with distance smaller than 5\,kpc (ref.\citep{Li+23}).
For future {\it Gaia} DR5, the volume will be significantly expanded with distance as far as 10\,kpc.
We here attempt to estimate the number of J0731+3717 like star ejected from globular cluster via Hills mechanism in current and future searching volume.
Doing so, we calculate mean ejection rate for 145 well known globular clusters in the full loss-cone and empty loss-cone regimes\citep{Subr19}:
\begin{equation}
\mathcal{R_{\rm full}} = f_{\rm b} \left(\frac{a}{0.1 {\rm AU}}\right)\left(\frac{n}{10^5 {\rm pc^3}}\right)\left(\frac{M}{10^3 M_{\odot}}\right)^{4/3} {\rm Myr^{-1}}\text{,}
\end{equation}
and 
\begin{equation}
\mathcal{R_{\rm empty}} = f_{\rm b} \left(\frac{n}{10^5 {\rm pc^{-3}}}\right)^2\left(\frac{M}{10^3 M_{\odot}}\right)^{3}\left(\frac{\sigma}{10 {\rm km s^{-1}}}\right)^{-9} {\rm Myr^{-1}}\text{.}
\end{equation}
Here, we set cluster binary fraction $f_{\rm b} = 8.8$\% (ref.\citep{JB15}), and a constant binary separation of $0.05$\,AU.
The central stellar number density $n$ is estimated by 
the cluster central mass density and typical mass of a star in globular cluster. The former can be found at \url{https://people.smp.uq.edu.au/HolgerBaumgardt/globular/parameter.html} and the latter is assumed to be 0.8\,$M_{\odot}$.
$M$ is the mass of central black hole which is extrapolated from $M$-$\sigma$ relation\citep{Gultekin09}.
$\sigma$ is the cluster central velocity dispersion that can again be found at \url{https://people.smp.uq.edu.au/HolgerBaumgardt/globular/parameter.html}.
The number of recent ejections for each cluster can thus be calculated by $N_{\rm ej} = \overline{\mathcal{R}}\Delta T$  in the past 250 Myr (i.e. $\Delta T = 250$\,Myr).
$\overline{\mathcal{R}}$ is the mean of $\mathcal{R_{\rm full}}$ and $\mathcal{R_{\rm empty}}$.
On average, the ejection rate is 1.34\,Myr$^{-1}$ for a cluster.
We then run MC simulations to perform the ejection experiments. 
In each simulation, all the orbits of 145 globular cluster are integrated back to 250 Myr with a time resolution of 0.1\,Myr, $N_{\rm ej}$ ejection events for each cluster are thus uniformly distributed in past 250 Myr.
For each event, a star is assumed to be kicked off from the cluster isotropically by a velocity $v_{\rm ej}$, which can be calculated from Equations 7 and 8 by assuming a binary containing two J0731+3717 like stars with a constant separation of 0.05\,AU.
All these ejected stars are then integrated to present-day.
In total,  we performed 100 MC simulations.
On average, over thirty thousand J0731+3717 like high-velocity stars with $V_{\rm GSR} \ge 400$\,km\,s$^{-1}$ are kicked off from clusters, corresponding to an ejection rate of around 10$^{-4}$\,yr$^{-1}$, which is well consistent with that estimated from a dynamical numerical simulation ( ref.\citep{FG19}).
As a comparison, only 10 such high-velocity stars, ejected in the past 250\,Myr are found in the Monte Carlo
{\it N}-body simulations (\url{https://zenodo.org/record/7599871}) for single-binary interaction involving compact objects by ref.\citep{CR23}.
Its rate, about $4 \times 10^{-8}$\,yr$^{-1}$, is three orders of magnitude lower than that of Hills mechanism we consider here.
In our simulations, the number of J0731+3717 like stars within 5 and 10\,kpc from the Sun are found to be $49^{+7}_{-7}$ and $480^{+48}_{-32}$, respectively.
The values and uncertainties are given by the PDF generated by 100 MC simulations.
If we consider the past 14 Gyr (the age of universe) instead of the above 250\,Myr, the number of ejected stars within 5\,kpc from the Sun is  $537_{-27}^{+24}$ (without considerations of dynamical effects).
%This number is comparable to the number of discovered extreme velocity stars (over 500 J0731+3717 like stars with $V_{\rm GSR} \ge 420$\,km\,s$^{-1}$ in our sample), if considering the unaccounted number ejected from dwarf galaxies.
This result shows that the current searching volume potentially contains a large number of high-velocity stars ejected from globular clusters but their orbits are difficult to be linked to their host clusters since most of them are not ejected recently. 
 We note that all the above simulation results are based on a series of parameters from previous studies, which may still be under debate. However, the predicted order of magnitude is meaningful.

Currently, we have found one of the nearly hundred cluster ejected high-velocity stars as predicted by our above MC simulations.
We expect dozens of more such star(s) can be discovered in the near future, with the fast increasing sampling rate of large-scale spectroscopic surveys (e.g., the LAMOST, SDSS-V and DESI). 
With the final release of {\it Gaia} DR5, the number can be even increased to a factor of ten, more details about the IMBHs hosted by globular clusters can be explored at that time.

\begin{table*}[htb]
\small{\textbf{Table S1}: Top 5 globular clusters sorted by the ratio of the closest orbital distance to the clusters' tidal radius (from small to large) in the backward orbital analysis of J0731+3717.}
\begin{center}
\begin{tabular}{ccccc}
\hline
Cluster name & Closest distance ($d_{\rm min}$) & Backward time & Tidal radius ($r_{t}$) & $d_{\rm{min}}/r_{\rm t}$\\
 & (kpc) & (Myr) & (kpc) & \\
\hline
M15 & 0.058 & 21.1& 0.132 & 0.44 \\
Pal 10 & 2.779 & 16.3 & 0.063 & 43.85 \\ 
NGC 6715 &12.867&  46.9 & 0.279 & 46.09\\
NGC 7089 & 5.813 & 17.5 & 0.111 & 52.47 \\
NGC 6121 & 2.845 & 1.7 & 0.054 &52.74\\
\hline
\end{tabular}
\end{center}
\end{table*}

\begin{table*}[htb]
\small{\textbf{Table S2}: Backward orbital results under different assumptions.}
\begin{center}
\begin{tabular}{ccc}
\hline
Changed assumptions & Closest orbital distance & Backward time\\
&(pc)&(Myr)\\
\hline
{\tt BovyMWPotential2014} &$56.3$&$21.3$\\
($U_{\odot}, V_{\odot}, W_{\odot}$)\,=\,($11.10, 12.24, 7.25$)\,km\,s$^{-1}$&$57.8$&$21.1$\\
$V_{c} (R_0) = 234.04$\,km\,s$^{-1}$&$59.1$&$21.1$\\
$R_0 = 8.34$\,kpc&$53.1$&$21.1$\\
 Modelling M15 as a moving Plummer potential&  55.6 & 21.1\\
Adding potential from LMC&  50.9 &  21.1\\
\hline
\end{tabular}
\end{center}
\end{table*}

\begin{figure*}
 \begin{center}
 \centerline{\includegraphics[scale=1.25,angle=0]{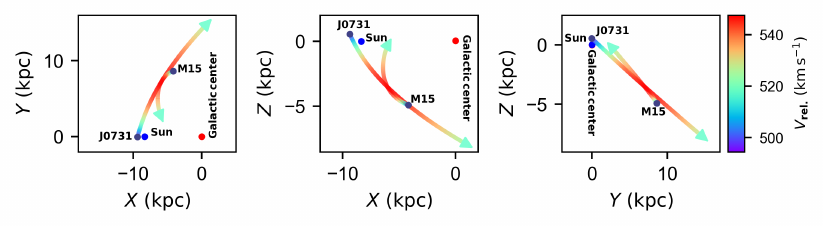}}
  \small{\textbf{Figure S1}: The 3D backward orbits of J0731+3717 and M15 projected in X–Y (left), X–Z (middle) and Y –Z (right) planes,
color coded by the relative velocity between each other as indicated by the right colorbar. The positions of the Galactic center and
the Sun are represented by the red and blue dots, respectively.}
\end{center}
\end{figure*}

\begin{figure*}
 \begin{center}
 \centerline{\includegraphics[scale=0.9,angle=0]{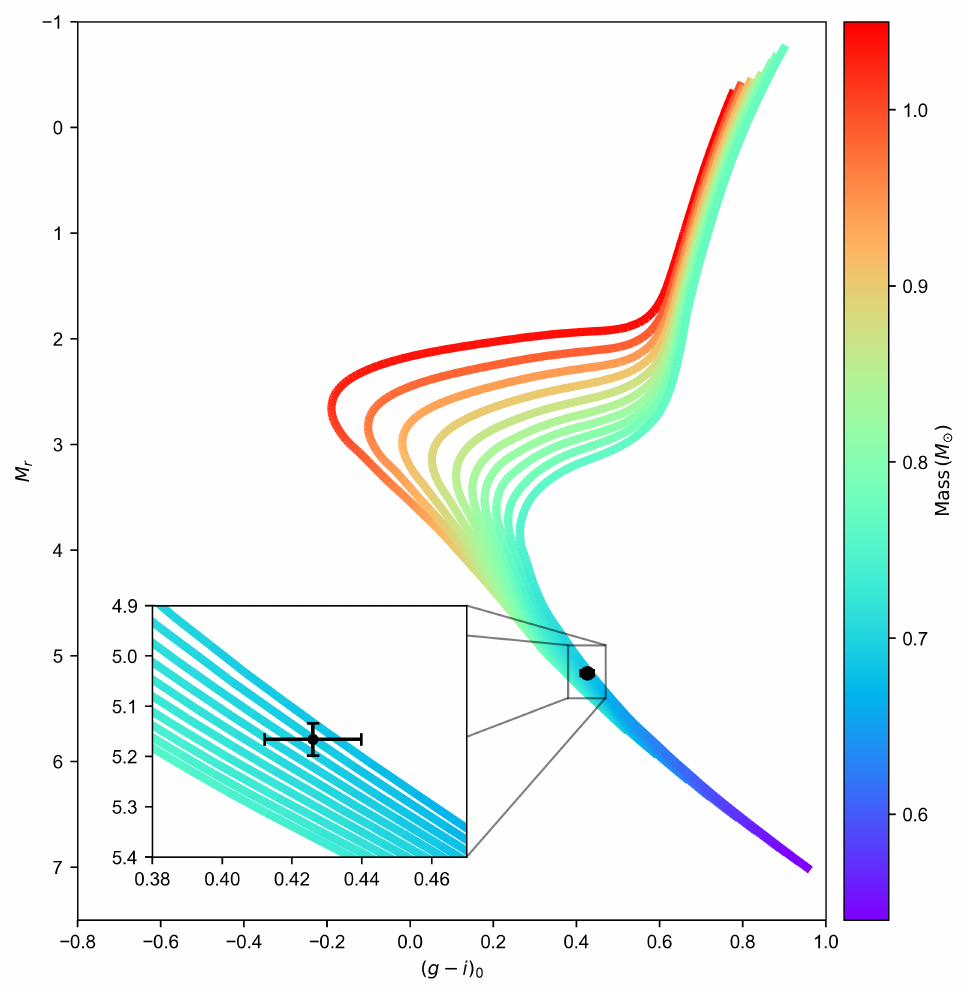}}
  \small{\textbf{Figure S2}: Comparison of J0731+3717 and stellar isochrones on $(g - i)_0$ versus $M_r$ diagram. The black dot with error bar marks the position of J0731+3717.
 The background shows ten isochrones with ages ranging from 5 to 14 Gyr in a step of 1 Gyr (left to right) taken from DSEP.
 The colors represent the stellar mass, as indicated by the right color bar.
 All the isochrones have constant [Fe/H] of $-2.23$ and [$\alpha$/Fe] of $+0.20$.
 Zoom inset shows the comparison more clearly.}
  \end{center}
\end{figure*}

\begin{figure*}
 \begin{center}
 \centerline{\includegraphics[scale=0.85,angle=0]{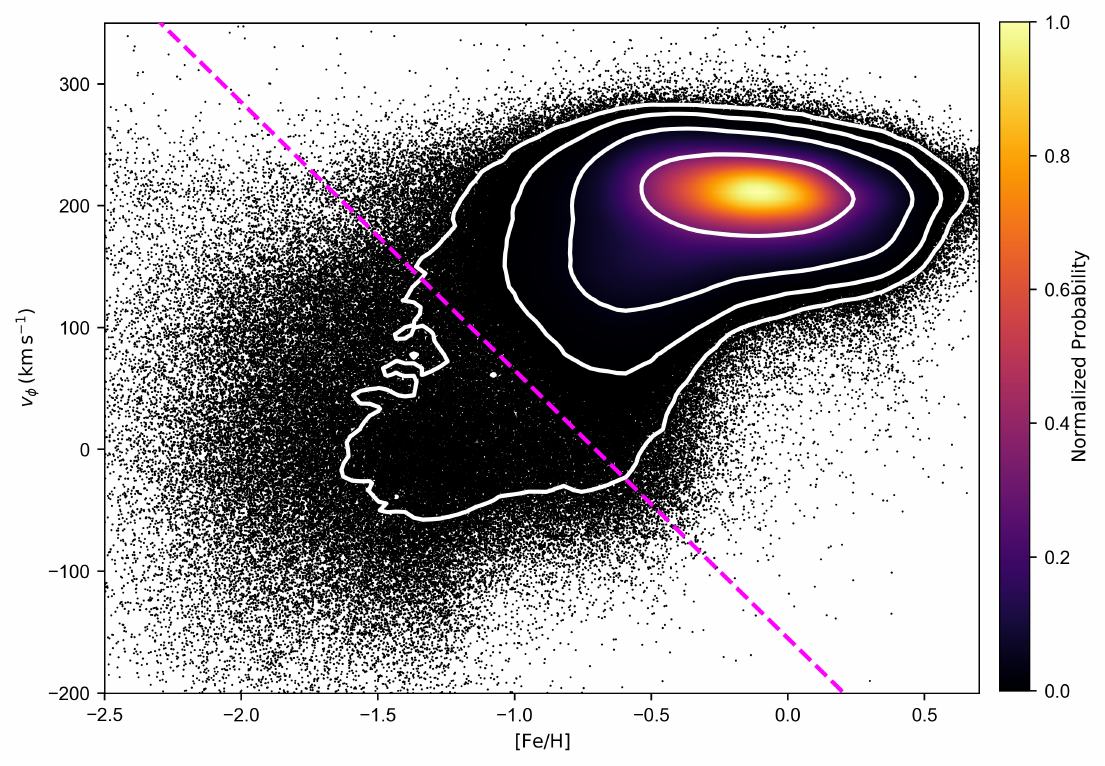}}
  \small{\textbf{Figure S3:}  The density distribution of $v_{\phi}$ vs. [Fe/H] for 5,020,788 stars observed by  SDSS DR12, LAMOST DR8, APOGEE DR16, and GALAH DR2 with spectral signal-to-noise ratio greater than 10. Similar to ref.\citep{Li+23}, all stars are required to have parallax greater than 0.2\,mas, parallax measurement uncertainty better than 20\% and RUWE smaller than 1.4. The magenta dashed-line is an empirical cut to separate the disk population (upper right) and halo population (lower left).}
\end{center}
\end{figure*}

\begin{figure*}
 \begin{center}
 \centerline{\includegraphics[scale=0.95,angle=0]{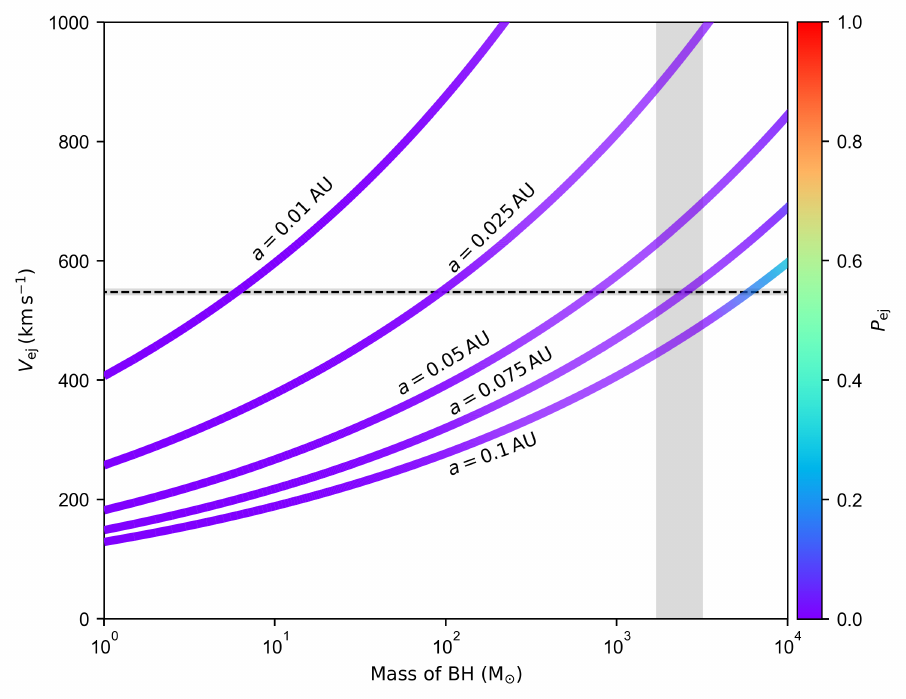}}
  \small{\textbf{Figure S4:} Ejection velocities predicted by Hills mechanism. This plot resembles Fig.~3, but with the closest distance between the binary and the black hole set to 3 AU. The ejection probability, as indicated by the colorbar on the right side, is almost zero for most black hole masses and binary separation configurations.}
\end{center}
\end{figure*}

%\printbibliography
%\end{refsection}